\documentclass[twoside]{ilcws08}
\usepackage[latin1]{inputenc}
\usepackage[dvips]{graphicx,epsfig,color}
\usepackage{wrapfig,rotating}
\usepackage{amssymb,amsmath,array}

\begin{document}
\title{Compton Cherenkov Detector Development for ILC Polarimetry}
\author{Christoph Bartels$^{1,2}$, 
  Christian Helebrant$^{1,2}$, 
  Daniela K\"afer$^1$, and 
  Jenny List$^1$
  \thanks{The authors acknowledge the support by DFG Li 1560/1-1.}
  \vspace{.3cm}\\
  1- Deutsches Elektronen Synchrotron (DESY) - Hamburg, Germany \\[0.4mm]
  2- Universit\"at Hamburg, Germany          \\[-1.4mm]
}

\definecolor{dred}       {rgb}{0.76, 0.00, 0.00}  \newcommand{\dred}       {\color{dred}}


\maketitle

\begin{abstract}
  In order to fully exploit the physics potential of the ILC, 
  it will be necessary to measure (and control) beam parameters to 
  a permille level precision. In case of the beam polarisation, this 
  can only be achieved with dedicated high energy Compton polarimeters 
  and by improving the detector linearity, as well as the calibration 
  of the analyzing power. 
  This note summarises results of an early testbeam period with    
  the Cherenkov detector of the SLD polarimeter,                   
  linearity measurements of readout electronics and photodetectors 
  and compares simulation results of the SLD Cherenkov detector    
  with those of a new `U-shaped' prototype.                    
\end{abstract}

\section{Polarisation and precision}
The measurement and control of beam parameters to a permille level precision 
will play an important role in the ILC's ambitious physics programme~\cite{bib:RDR,bib:POWER}. 
This means, not only the luminosity and the beam energy need to be measured precisely, 
but also the polarisation of the electron and positron beams have to determined with 
unprecedented accuracy.
While this has already been achieved at previous colliders for beam energy 
measurements, the precision of polarisation measurements has to be improved by 
at least a factor of two compared to the up to now most precise polarisation 
measurement of the SLD polarimeter~\cite{bib:SLDpolmeas}. 
It is planned to achieve $d{\mathcal P}/{\mathcal P}=0.25$\% or better. 

The polarisation measurement at the ILC will combine the measurements of two 
dedicated Compton polarimeters, located upstream and downstream of the $e^+e^-$ 
interaction point, and data from the $e^+e^-$ annihilations themselves. 
While $e^+e^-$ annihilation data will finally (after several months) provide an 
absolute scale, the polarimeters allow for fast measurements and, in case of the 
upstream polarimeter, probably even resolve intra-train variations, give feedback 
to the machine, reduce systematic uncertainties and add redundancy to the entire 
system~\cite{bib:exsum-epws08}. 

Circularly polarised laser light hits the $e^+$($e^-$)-beam under a small angle 
and typically in the order of 1000 electrons are scattered per bunch. The energy 
spectrum of the scattered particles depends on the product of laser and beam 
polarisations, so that the measured rate asymmetry w.r.t. the (known) laser helicity 
is directly proportional to the beam polarisation. 
Since the electrons' scattering angle in the laboratory frame is less than 10~$\mu$rad, 
a magnetic chicane is used to transform the energy spectrum into a spacial distribution 
and lead the electrons to the polarimeter's Cherenkov detector. 
It consists of staggered `U-shaped' aluminum tubes lining the tapered exit window 
of the beam pipe. The tubes are filled with the Cherenkov gas C$_{\rm 4}$F$_{\rm 10}$ 
and are read out by photodetectors. Electrons traversing the basis of these 
`U-shaped' tubes generate Cherenkov radiation which is reflected upwards to 
the photodetectors~\cite{bib:kaefer-epws08}.

Developing a Cherenkov detector suitable for achieving the aforementioned 
precision of $d{\mathcal P}/{\mathcal P}=0.25$\% demands improvements in 
various areas of the experimental setup. 
Of utmost importance, however, is the linearity of the detector response, 
or the ability to control and correct for a non-linear response. 
The following areas of work are briefly discussed: testbeam data analysis, 
linearity measurements of readout electronics and photodetectors (PDs), 
simulation and design of a new prototype detector.

\section{Testbeam results}
\begin{figure}[!h]
  \setlength{\unitlength}{1.0cm}
  \begin{picture}(10.0, 4.30)
    \put(-0.1,-0.2)  {\epsfig{file=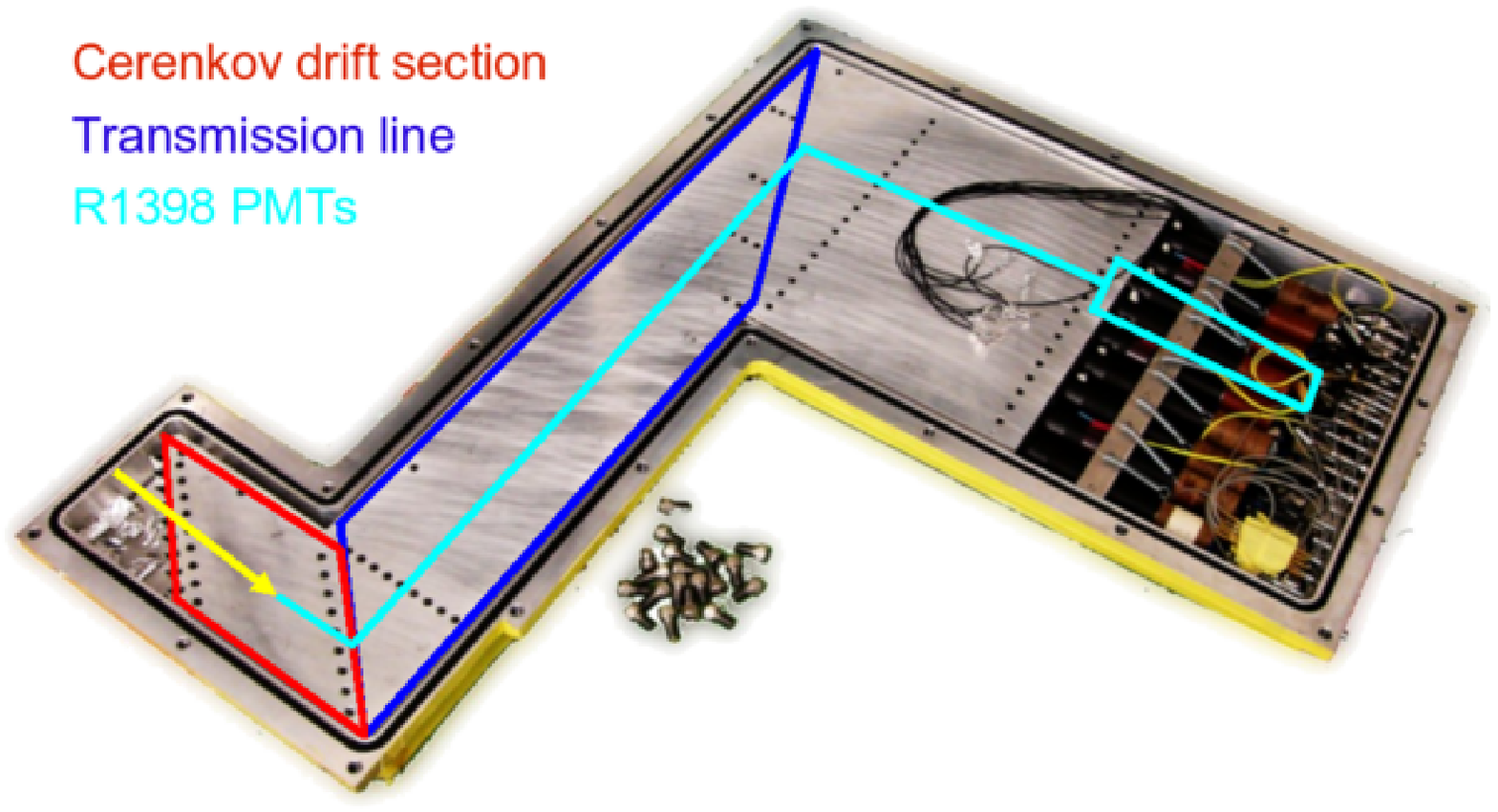, clip= , angle=0,   width=0.50\linewidth}}
    \put( 7.0, 4.8)  {\epsfig{file=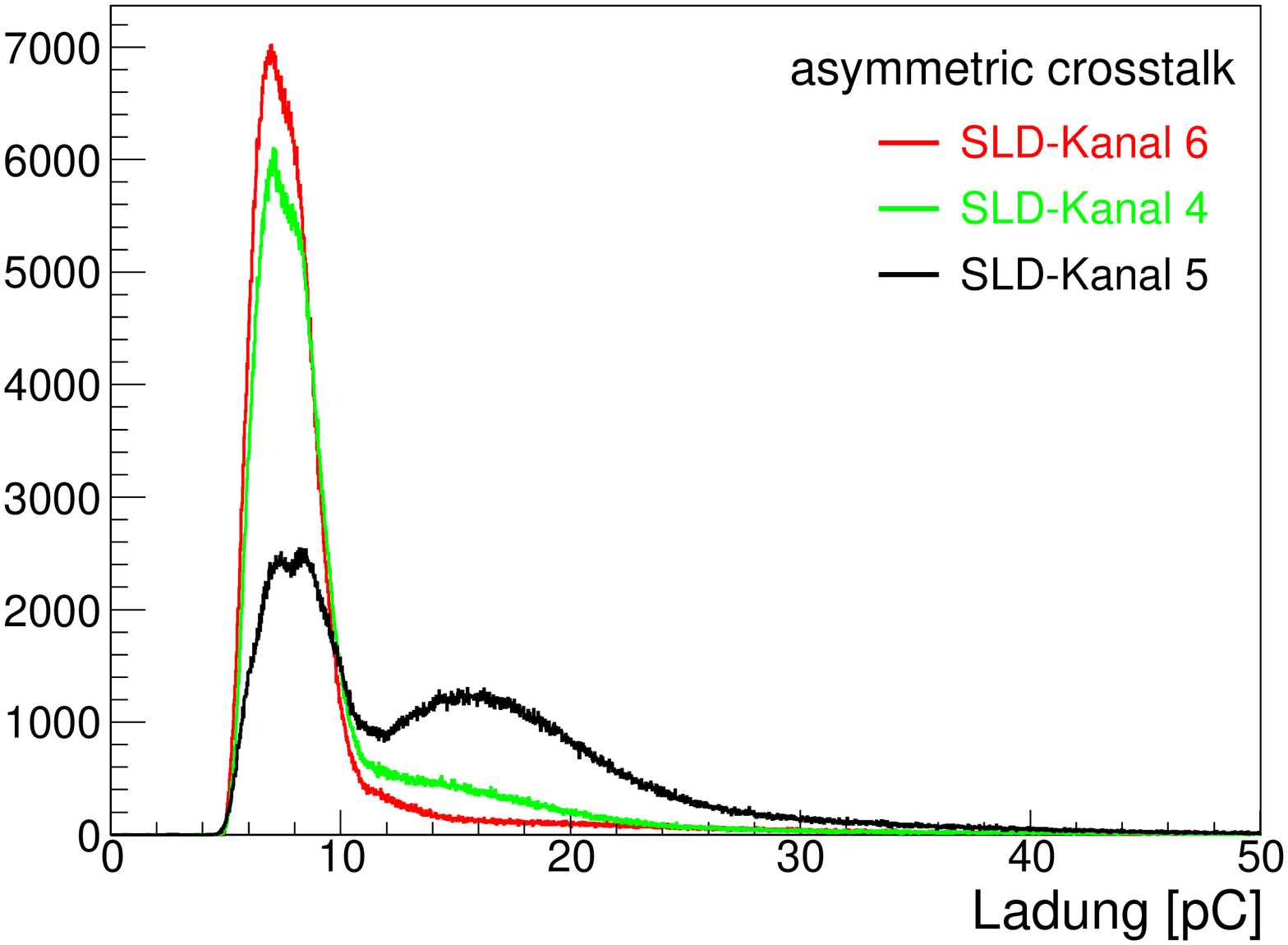, clip= , angle=-90, width=0.50\linewidth}}
    \put( 0.1, 4.4)  {(a)}
    \put( 6.5, 4.4)  {(b)}
  \end{picture}
  \caption{(a) The open SLD Cherenkov detector; 
    different colours indicate different sections.
    (b) Asymmetric cross talk in channels~4 and~6, 
    while the beam is centered on channel~5.}
  \label{fig:SLD-det-crosstalk}
\end{figure}

In November and December 2007, the SLD Cherenkov detector 
(Fig.~\ref{fig:SLD-det-crosstalk}(a)) was set up in the DESY-II testbeam for two 
short periods of time. 
From different orientations of the detector, either using the Cherenkov drift 
section (indicated in red) or not (beam incident on the area indicated in blue), 
the reflectivity of the aluminum channels was measured to be about 90\%. 
As can be seen from Fig.~\ref{fig:SLD-det-crosstalk}(b), the cross talk between 
adjacent channels is asymmetric with more Cherenkov photons detected in channels 
left of the one on which the beam is centered. 

Further tests included a setup where some of the original photodetectors were 
exchanged for newer types, especially three multianode photomultipliers and also 
a new silicon-based photomultiplier. However, results from these measurements 
are not discussed here.

\section{Linearity measurements of electronics and photodetectors}
A component test stand is used to study different types of photodetectors. 
They are set up in a light-tight box and read out via a high resolution double range 
12-bit VME charge-to-digital converter (QDC), with either 200~fC or 25~fC per LSB 
(least significant bit).
A blue LED ($\lambda =\;$472~nm) connected to a function generator is used to 
generate the light detected by the PD. Before testing the linearity of the PDs 
themselves, the differential and integral non-linearity (DNL, INL) of the QDC 
has also been measured. 

A ramp ($f =\;$10~Hz) is used as input signal while the QDC is triggered by 
a short random gate of 50~ns duration. 
The probability for each transition to occur at a certain QDC code bin is 
measured and compared to an ideal distribution. 
If the QDC was ideal, a uniform distibution of code bins would be expected. 
The ratio between the measured and the ideal distribution is the code bin width, 
from which the DNL can be calculated as the deviation of the ideal code bin width 
of 1~LSB. The INL for a certain QDC code bin is then given by the sum of DNLs up 
to this bin. Both distributions are rather flat and a fit to the mid range of codes, 
from 200 to 800 QDC bins, gives an INL from 1 to 2 LSBs, corresponding to 0.1-0.2\% 
of the full scale range. 

\begin{figure}[!h]
  \setlength{\unitlength}{1.0cm}
  \begin{picture}(14.0, 9.08)
    \put( 0.00, 4.50) {\epsfig{file=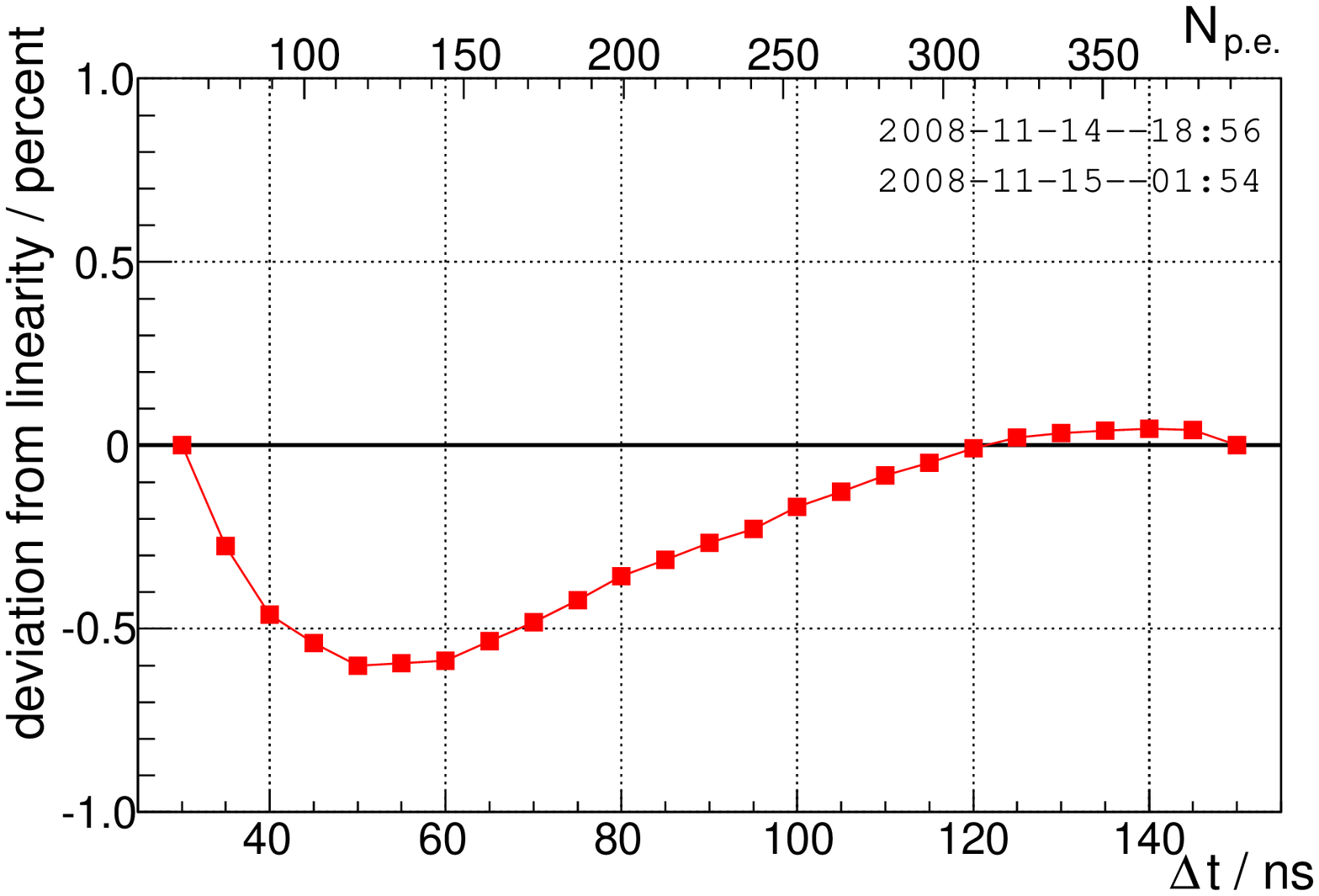, clip= , width=0.50\linewidth}}
    \put( 7.10, 4.50) {\epsfig{file=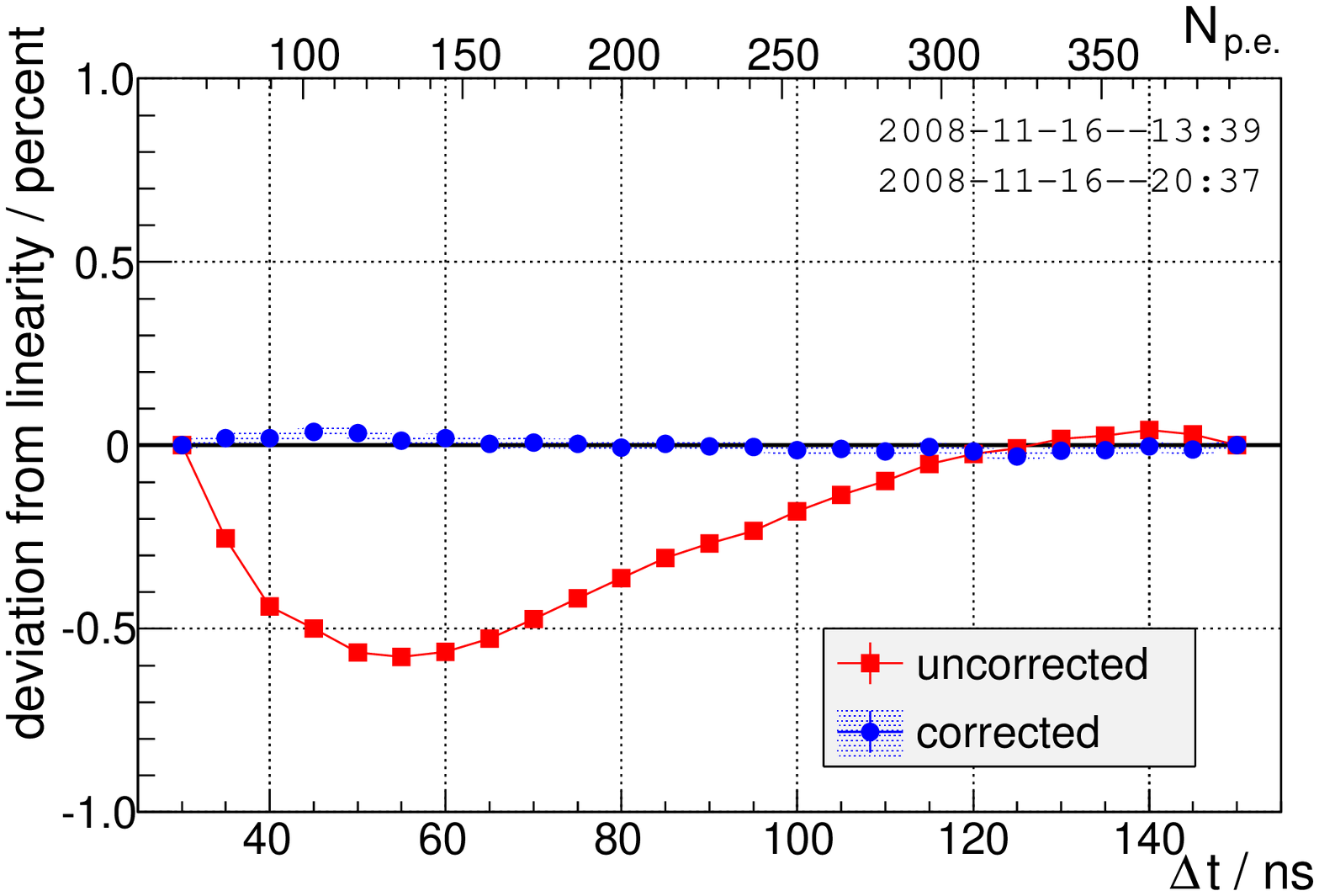, clip= , width=0.50\linewidth}}
    \put( 0.00,-0.35) {\epsfig{file=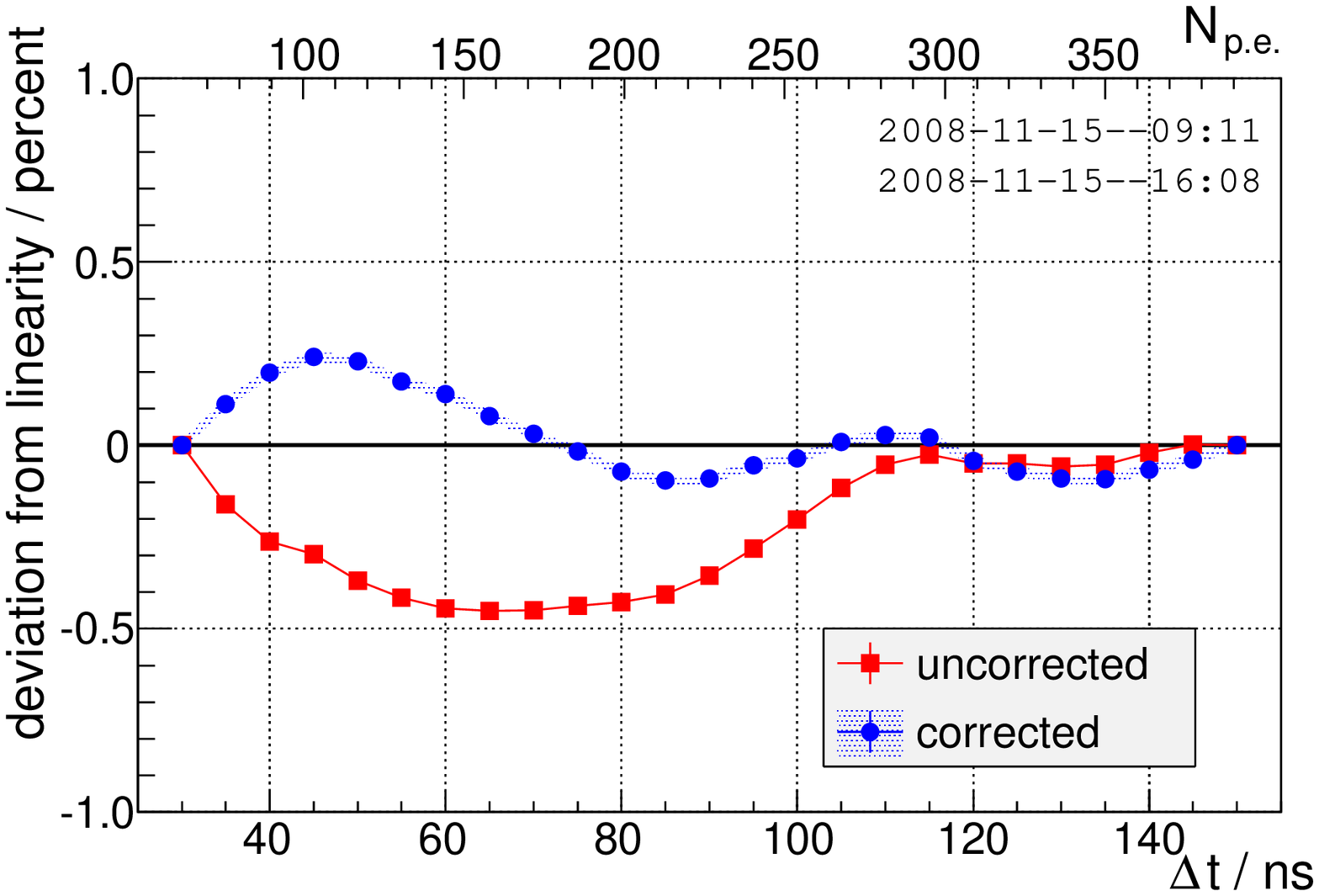, clip= , width=0.50\linewidth}}
    \put( 7.10,-0.35) {\epsfig{file=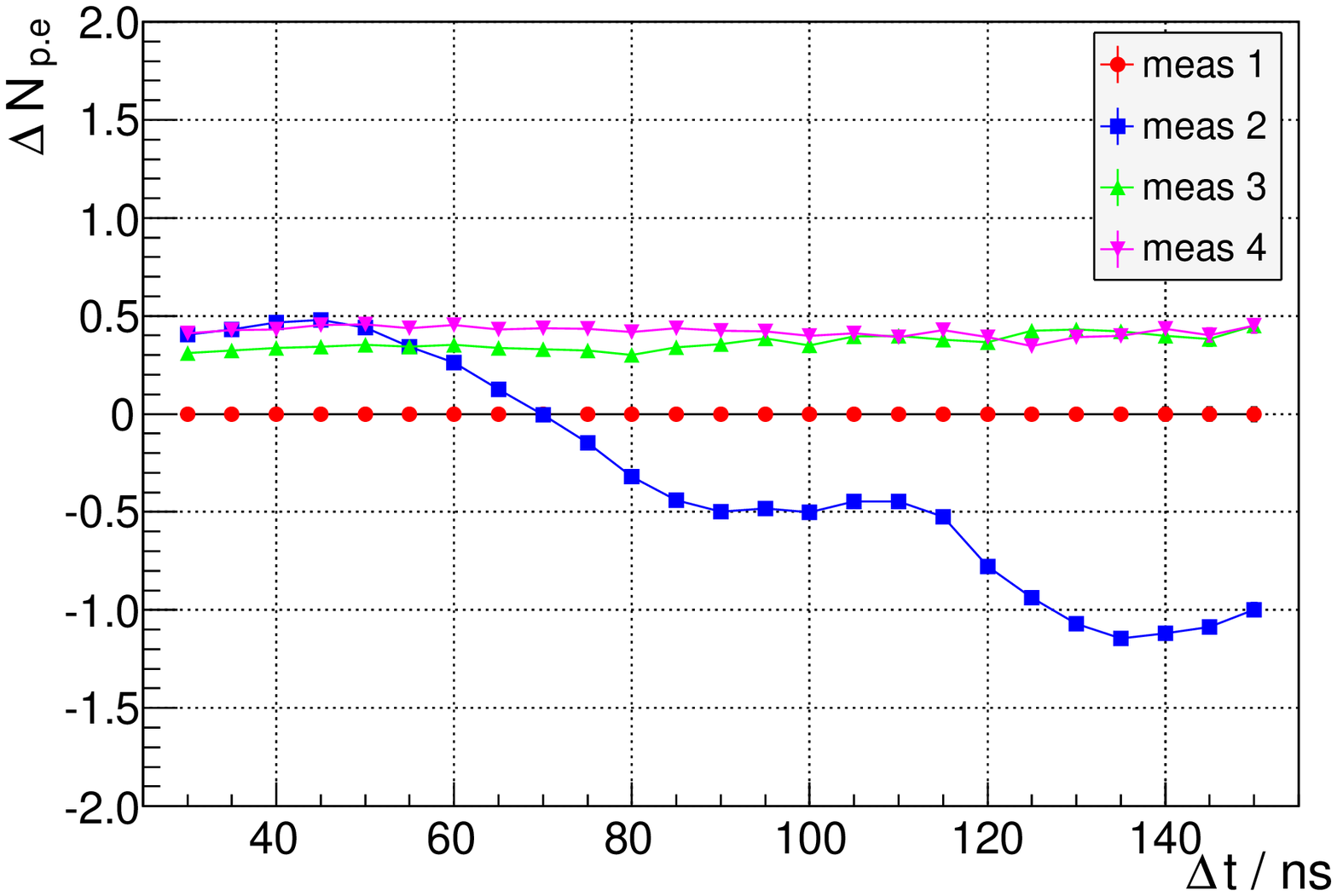, clip= , width=0.50\linewidth}}
    \put( 0.95, 5.20) {(a)}
    \put( 8.05, 5.20) {(b)}
    \put( 0.95, 0.35) {(c)}
    \put( 8.05, 0.35) {(d)}
    \put( 1.00, 8.00) {\small 1$^{st}$ measurement}
    \put( 1.45, 7.60) {\small $\to$ reference}
    \put( 8.10, 8.00) {\small 3$^{rd}$ \& 4$^{th}$ series similar}
    \put( 8.10, 7.60) {\small corr. successful: NL$\,\ll\,$0.1\%}
    \put( 1.00, 3.55) {\small 2$^{nd}$ series}
    \put( 1.00, 3.15) {\small corr. fails: NL$\approx\,$0.25\%}
    \put( 8.10, 3.80) {\small time ordering}
  \end{picture}
  \caption{PD linearity measurements obtained using the pulse-length method: 
    (a) reference measurement, (b) 2$^{nd}$ and (c) 3$^{rd}$ measurement series, 
    corrected using the 1$^{st}$ measurement.
    (d) The four measurement series ordered in time.}
  \label{fig:PD-linearity}
\end{figure}

The most extensive linearity studies have been performed on 2x2 multianode 
photomultipliers (MAPM, Hamamatsu R5900U-M4). The spectrum of QDC counts is fitted 
by a modified Poisson function to determine the number of incident photoelectrons. 
The `true' number of photoelectrons expected for a certain amount of light cannot 
be determined since the relation between light yield and LED bias voltage is not 
calibrated. 
However, the method used to measure the PD's INL is independent of the absolute 
scale of the LED and only depends on the length of a rectangular pulse lighting 
the LED. Varying this pulse length ensures a linear variation of the amount 
of light on the photocathode of the PD. The pulse length is varied between 30~ns 
and 150~ns in steps of 5~ns. 
\enlargethispage{1mm}
Figure~\ref{fig:PD-linearity} shows the results of several measurement series. 
As can be seen in Fig.~\ref{fig:PD-linearity}(b), the correction is highly 
successful for the 3$^{rd}$ and 4$^{th}$ measurement series with a resulting 
INL of less than 0.1\%, but fails for the 2$^{nd}$ measurement 
(Fig.~\ref{fig:PD-linearity}(c)), where the INL is only about 0.25\%. 
However, these results show that the PD non-linearity can be measured and 
thus controlled to a precision of 0.1\%.
A second method relying on optical filters will be used to cross check 
these results.

Another two methods were developed to measure the DNL: one measures the 
difference between the PD's response to the initial (variable) pulse $P_i$ 
and its response to the pulse $P_i + p$ with $p$ being a very small fix pulse. 
For the second method, a 4-hole mask is applied to the PD so that the ratio 
of the sum of single pulses through each hole and one pulse through all four 
holes simultaneously gives the DNL.

\section{Prototype simulation and design}
A smaller prototype has been simulated and constructed to study the entire Cherenkov 
detector design. This simplified version of the envisioned detector consists of only 
two U-shaped aluminum channels with a cross section of 8.5~mm$^2$ embedded in a box 
flooded with the Cherenkov gas C$_4$F$_{10}$. The gas is non-flammable and was chosen 
mainly because of its high Cherenkov threshold of about 10~MeV. 
The electron beam will pass through the basis of the U-shaped channels, entering and 
exiting through thin aluminum windows, and produce Cherenkov photons which are then 
reflected towards a photodetector mounted on the hind U-leg. The front U-leg is solely 
for calibration purposes with two LEDs mounted there, one per channel. 

\begin{figure}[!h]
  \setlength{\unitlength}{1.0cm}
  \begin{picture}(10.0, 4.2)
    \put( 0.20, 4.3)  {\epsfig{file=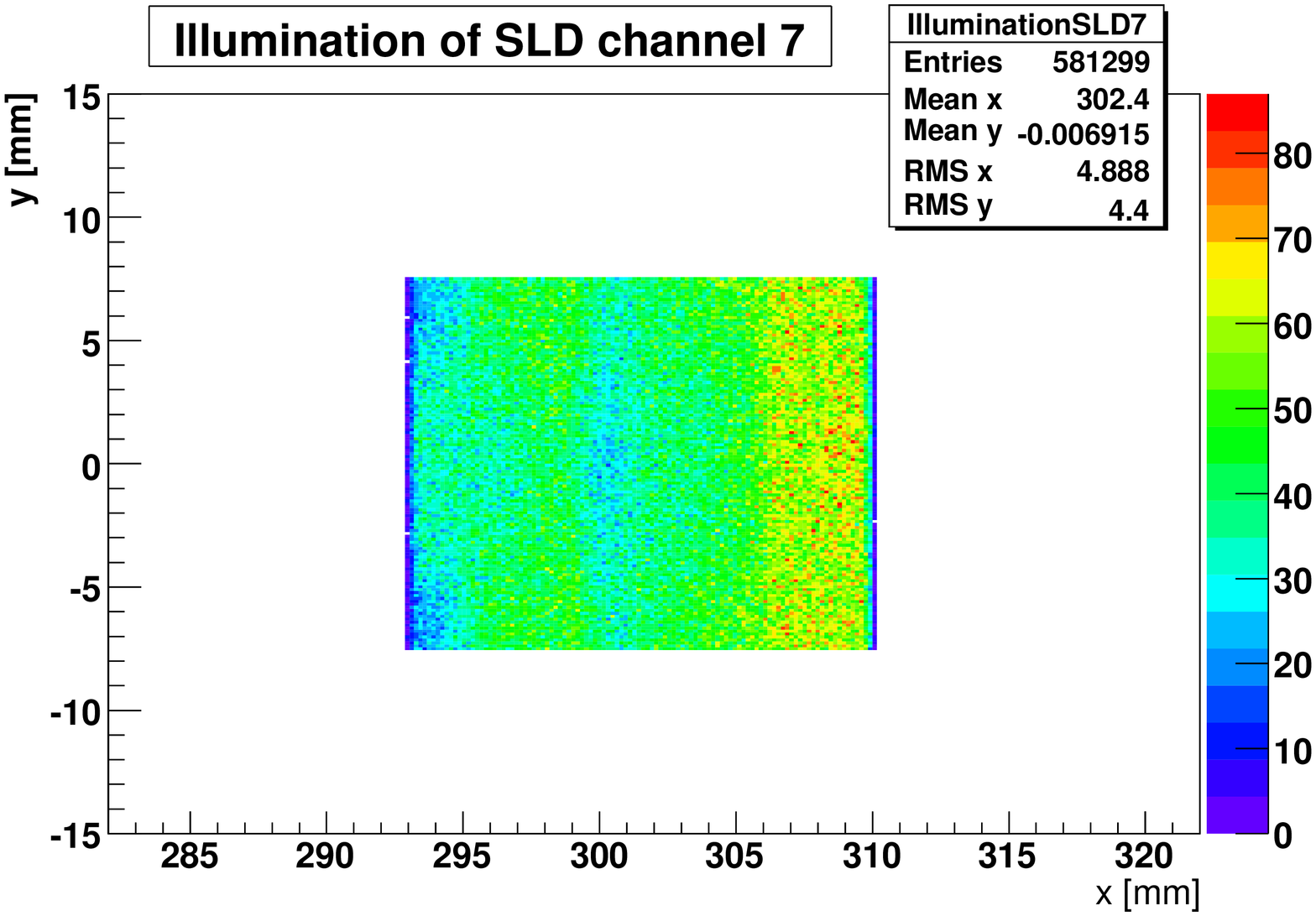, clip= , angle=-90, width=0.47\linewidth}}
    \put( 7.30, 4.3)  {\epsfig{file=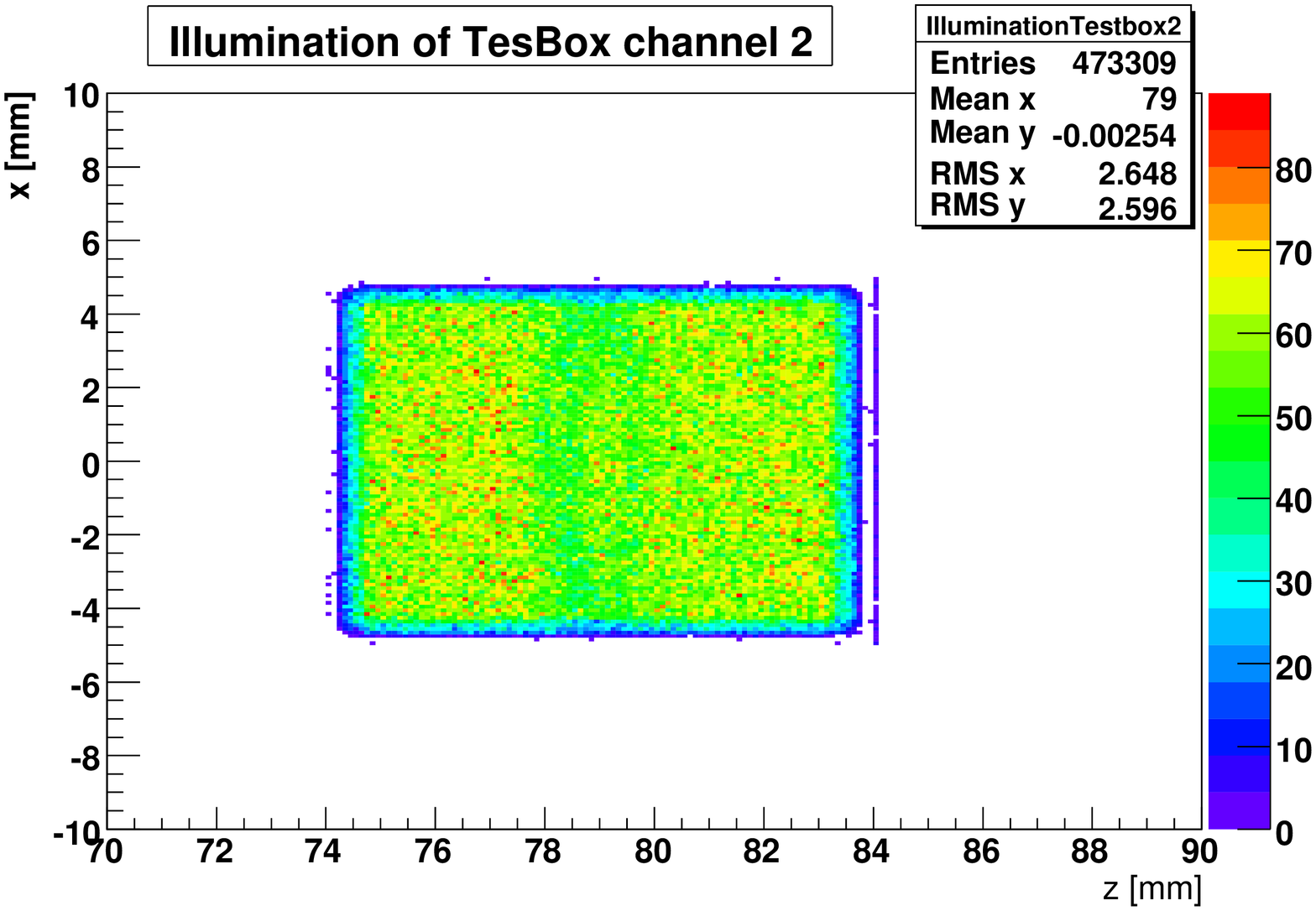, clip= , angle=-90, width=0.47\linewidth}}
    \put( 0.02,-0.2)  {(a)}
    \put( 7.10,-0.2)  {(b)}
  \end{picture}
  \caption{Simulation of the illumination (light intensity) in a single 
    channel for (a) the SLD Cherenkov detector and (b) the new prototype.}
  \label{fig:illu-SLD-testbox}
\end{figure}

For the design of the prototype detector and also for the interpretation of future 
testbeam data, an optical simulation based on \texttt{GEANT4} has been created. 
However, the PD response, i.e. mainly the PD's quantum efficiency is simulated 
separately using simple ROOT macros. 
Figure~\ref{fig:illu-SLD-testbox} shows a comparison between the simulated 
illumination of a single channel of the SLD detector and the new U-shaped prototype. 
The strong asymmetry visible for the SLD-type channel in 
Fig.~\ref{fig:illu-SLD-testbox}(a) is avoided with the new design 
(Fig.~\ref{fig:illu-SLD-testbox}(b)), where only a slight inhomogeneity 
in the light intensity is seen. The reduction of cross talk and the avoidance 
of geometry-based asymmetries was the main reason to choose the peculiar 
U-shaped geometry for the prototype detector. 

\begin{figure}[!h]
  \setlength{\unitlength}{1.0cm}
  \begin{picture}(10.0, 8.1)
    \put( 0.20, 4.7)  {\epsfig{file=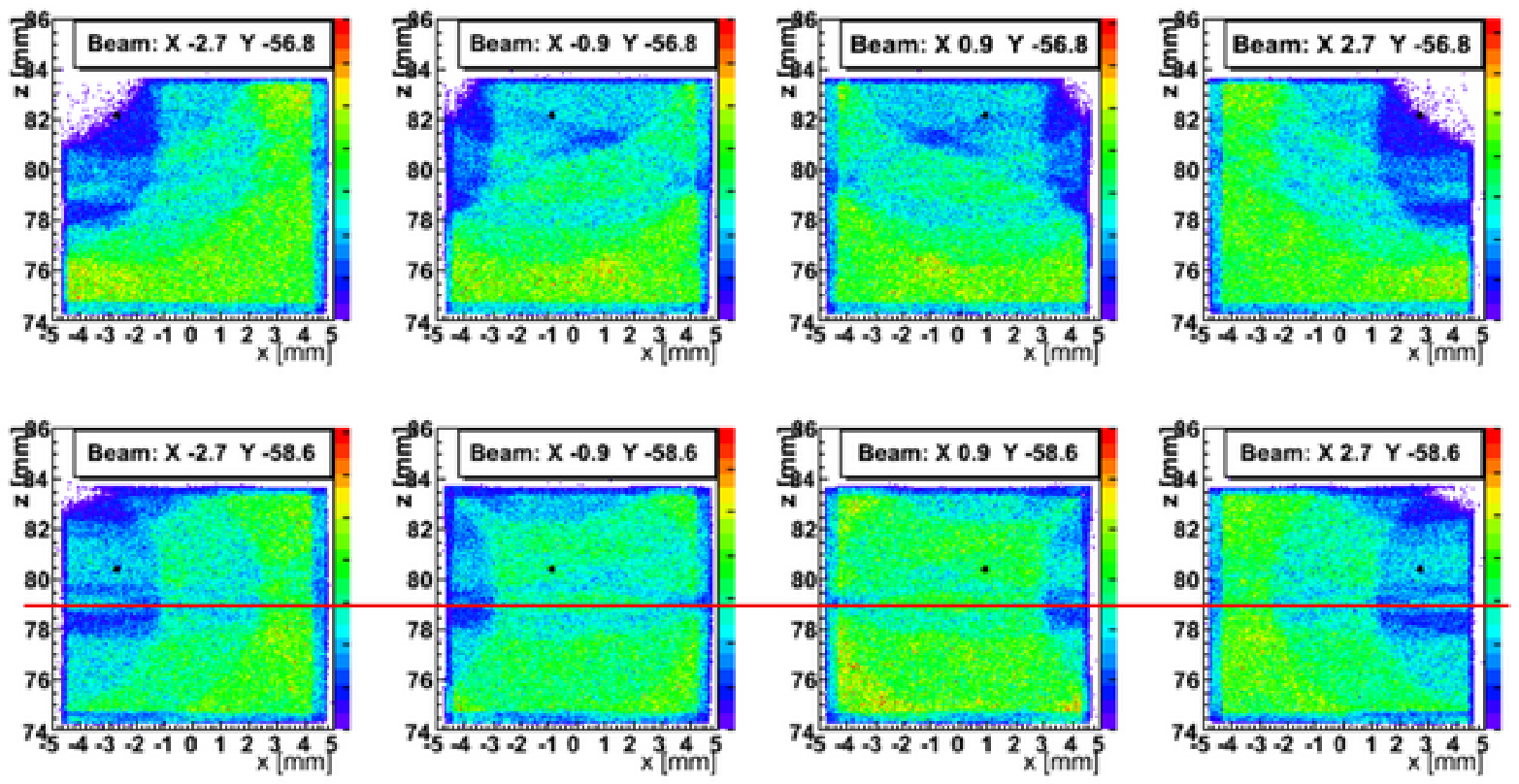, clip= , width=0.98\linewidth}}
    \put( 0.20,-0.2)  {\epsfig{file=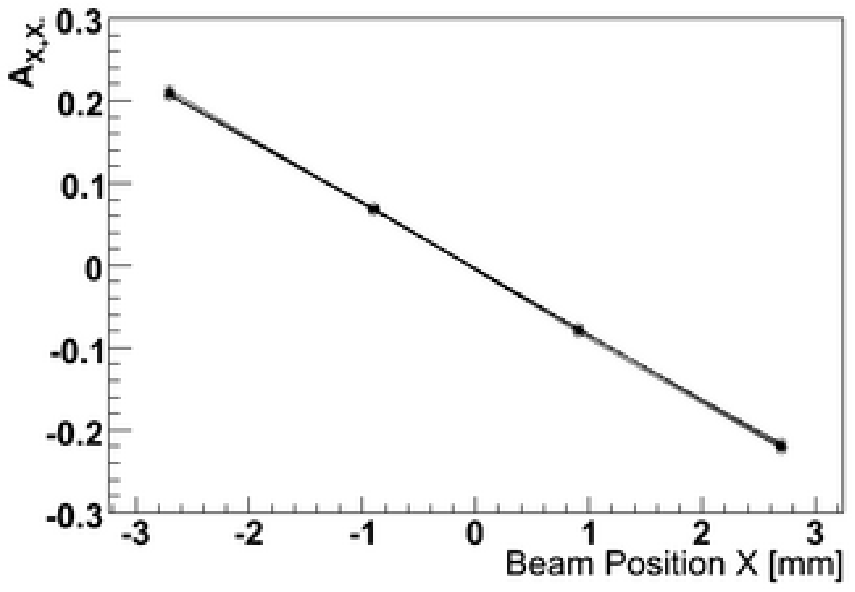, clip= , width=0.47\linewidth}}
    \put( 7.40,-0.2)  {\epsfig{file=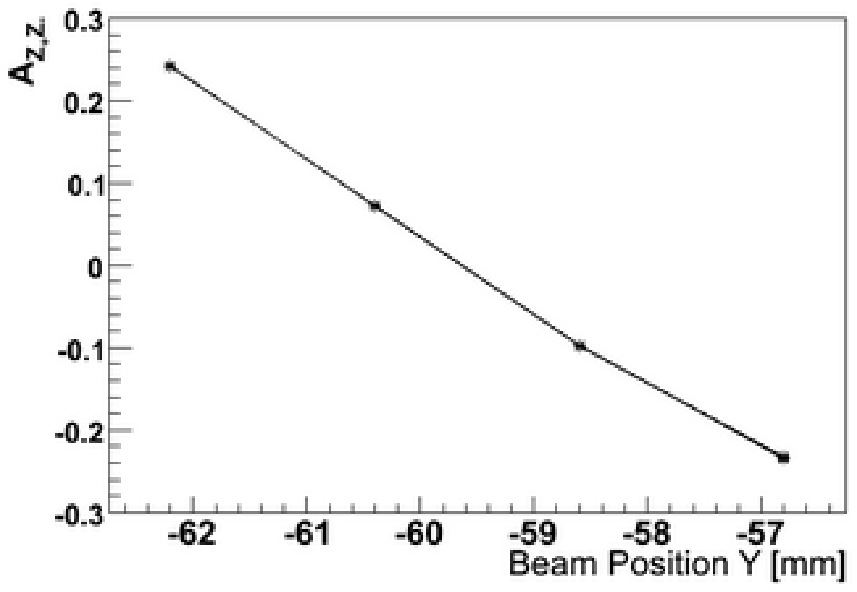, clip= , width=0.47\linewidth}}
    \put( 0.01, 4.5)  {(a)}
    \put( 0.02,-0.1)  {(b)}
    \put( 7.20,-0.1)  {(c)}
    \put( 3.50, 3.8)  {\small z fixed: 4 diff. graphs}
    \put( 3.50, 3.4)  {\small for 4 diff. x-positions}
    \put( 4.00, 3.0)  {\small\dred $\Rightarrow$ good linearity!}
    \put(10.60, 3.8)  {\small x fixed: 4 diff. graphs}
    \put(10.60, 3.4)  {\small for 4 diff. z-positions}
    \put(11.10, 3.0)  {\small\dred $\Rightarrow$ good linearity!}
  \end{picture}
  \caption{(a) Illumination scan of a testbox channel with 10,000 $e^-$ per shot. 
    (b,c) Asymmetries calculated from the light intensity on the 
    photocathode for different scans: $A_{x+x-}$ for fixed $z$ position
    and $A_{z+z-}$ for fixed $x$ position.}
  \label{fig:illu-Scan-Asym}
\end{figure}

An illumination scan with 4$\times$4 points per channel and 10,000 electrons per 
shot leads to Fig.~\ref{fig:illu-Scan-Asym}, where (a) depicts the light yield on 
the PD anode for electrons entering the Cherenkov section at a fixed $y$ position, 
but variable $x$ position. (Due to the geometry, a certain $z$ position on the PD 
andode corresponds to a certain beam position $y$ in the channels at the U-basis.)
Figure~\ref{fig:illu-Scan-Asym}(b,c) show the corresponding calculated asymmetries 
in the light intensity for scans in the $x$- and $z$-directions, respectively. 

The photodetector mounting on the hind U-leg is realized exchangeably to 
enable the testing of several different PDs (and different read-out modes) 
within one setup.


\begin{footnotesize}
%

\end{footnotesize}


\end{document}